%% file: main.tex
\documentclass[a4paper,reprint,aps,prl,amsfonts,amssymb,amsmath,nobibnotes,twoside,balancelastpage,nofootinbib,superscriptaddress]{revtex4-1}

\usepackage[danish,english]{babel}
\usepackage[T1]{fontenc}
\usepackage{tabularx} 
\usepackage{amsmath}  
\usepackage{graphicx} 

\usepackage{wrapfig}
\usepackage{gensymb}
\usepackage{mathtools}
\usepackage[utf8]{inputenc}
\usepackage{float}
\usepackage[breaklinks=true]{hyperref}
\usepackage{booktabs}
\usepackage{multirow}
\usepackage{comment}
\usepackage[Symbol]{upgreek}
\usepackage{color}
\usepackage{xcolor}
\usepackage{svg}
\usepackage[super]{nth}

\usepackage{tikz}

\begin{document}
\title{Effects of ground movements on realistic guide models for the European Spallation Source}
\input{Article.tex}

\end{document}

%% file: Article.tex
\setlength{\abovedisplayskip}{-3pt}

\title{Effects of ground movements on realistic guide models for the European Spallation Source}

\date{\today}
\author{Asla Husgard}
\author{Martin A. Olsen}
\author{Rebekka Fr\o ystad}
\author{Jonas P. Hyatt}
\affiliation{Nanoscience Center, Niels Bohr Institute, University of Copenhagen, DK-2100 Copenhagen}
\author{Mads Bertelsen}
\affiliation{Data Management and Software Centre, European Spallation Source ERIC, Copenhagen, Denmark}
\author{Rasmus Toft-Petersen}
\affiliation{Department of Physics, Technical University of Denmark, DK-2800 Kgs.\ Lyngby, Denmark}
\author{Kim Lefmann}
\affiliation{Nanoscience Center, Niels Bohr Institute, University of Copenhagen, DK-2100 Copenhagen}

\begin{abstract}
We model the effect of ground movement, based on empirical experience, on the transport properties of long neutron guides by ray-tracing simulations.  Our results reproduce the large losses found by an earlier study for a simple model, while for a more realistic engineering model of guide mounting, we find the losses to be significantly smaller than earlier predicted. A detailed study of the guide for the cold neutron spectrometer BIFROST at the European Spallation Source shows that the loss is 7.0(5) \% for wavelengths of $2.3-4.0\;$\AA; the typical operational wavelength range of the instrument.  This amount of loss does not call for mitigation by overillumination as suggested in the previous work. 
Our work serves to quantify the robustness of the transport properties of long neutron guides, in construction or planning at neutron facilities worldwide.
\\

\end{abstract}
\maketitle 

\section{Introduction}
Guide systems are indispensable at large-scale facilities for neutron scattering. The primary purpose of a guide is to transport the neutron beam far from the source, where the fast-neutron background is smaller \cite{Maier-Leibnitz1963}. In addition, the last part of the guide may converge to increase the neutron intensity \cite{Mildner1982} and potentially completely focus the beam to the sample under investigation \cite{Stahn2012,Boni2013,Weischelbaumer2015,Stahn2016,Hansen2017}. Furthermore, the longer source-sample distance will increase the neutron flight time, which can be of considerable advantage for time-of-flight instruments. This is particularly the case for long-pulse neutron sources with long guides \cite{Mezei1997,Mezei1997b,Schober2008}, such as the European Spallation Source (ESS), presently under construction \cite{Lefmann2013,Andersen2018,Andersen2020,ESS}, as well as the planned Second Target Station at the Spallation Neutron Source (SNS) \cite{SNSTS2}.
With the correct geometry and use of proper supermirrors, a ballistically shaped guide can potentially transport up to 80-90\% of the theoretical maximum of the desired neutron phase space from the moderator to the sample \cite{Schanzer2004,kleno2012,Cussen2013}.

The length of the neutron guides at 8 of the first 15 ESS instruments, including the inverse-geometry spectrometer BIFROST, will be more than 150~m \cite{Andersen2020}.
This is considerably longer than most guide systems previously built. For this reason, it is necessary to be particularly cautious with the design of these guides.

The weight of biological shielding needed in the construction of ESS is expected to cause the ground, and with that the foundation of the guide systems, to sink in an inhomogeneous fashion. As the guides are manufactured in pieces mounted on pillars, this will result in misalignment between pieces, in turn affecting the transport properties of the guides.

This issue has been investigated recently by Zendler {\em et al.}\ \cite{Zendler2016}. They consider two types of misalignment: those occurring due to the ground sinking under the shielding weight, modeled based on experience from SNS and referred to as systematic ground movements, and those caused by the uncertainty in the assembly of individual guide pieces, referred to as random misalignment. Using the Monte Carlo ray-tracing software VITESS \cite{VITESS1,VITESS2}, they showed that systematic ground movements of $6$~mm (maximum amplitude) and additional random displacement of $50~\upmu$m leads to a wavelength- and guide-quality-independent loss of 24.7(3)\% in neutron intensity for a 150 m long straight guide of cross section $10~\times~10$ cm$^2$. For a $3~\times~3$ cm$^2$ guide this loss increases to 59.8(5)\%. These losses are unacceptably high, leading to the suggestion of mitigation by overilluminating the guides. This does, however, in itself lead to a loss of guide performance, stressing the need for reinvestigating this problem.

In the work by Zendler {\em et al.}, the misalignments were implemented by shifting horizontal guide pieces by a vertical offset, resulting in gaps through which neutrons can escape \cite{Zendler2016}, as we will detail in the following. This geometry is, however, only a first-order approximation.

In the present work, we perform a simulation of the effect of ground movements on guide transmission. We implement a more realistic model, where the guide pieces will pivot around a mounting point close to the ends of the guide pieces, as this is consistent with the realistic guide geometry described by one vendor \cite{snag}.
We find that this pivot model leads to limited losses, of the order 5\% for a $3 \times 3$ cm$^2$ guide, much smaller than found by Zendler {\em et al}. For a realistic model of the BIFROST guide, we find that the total losses are 7.0(5)\%.
This more realistic model of guide motion therefore represents a less pessimistic scenario for the effect of ground movements and does not warrant the use of overillumination.


\section{Simulation method}
\subsection{Simulation of guides}
The interior of neutron guides are coated with neutron-reflecting mirrors. The size of the critical scattering vector $\mathbf{q_c}$ for such mirrors is given by

\begin{align}
q_c = \frac{4\pi}{\lambda} \sin (\theta_c(\lambda))
\end{align}
where $\theta_c$ is the critical angle of the mirrors and $\lambda$ is the neutron wavelength. We employ the default McStas value for the critical scattering vector for Ni, $q_{c,Ni} = 0.0219 ~$\AA$^{-1}$, corresponding to $\theta_c \approx 0.100^\textrm{o} \cdot \frac{\lambda}{\textrm{Å}}$. Neutrons with larger incoming angles are not totally reflected, and at a certain point the reflectivity drops rapidly. We can describe this dropping point by means of the $m$-value

\begin{align}
q_c = m ~q_{c,Ni}.
\end{align}
To describe the effective mirror reflectivity intermetiate $q$-values, we use the slope of the reflectivity profile $\alpha$,

\begin{align}
\alpha = \frac{\partial R}{\partial q}.
\end{align}
A plot of the reflectivity profile for different $m$-values as well as further information can be found in Ref. \cite{SwissOnline}. 

The efficiency of neutron guide systems is typically evaluated by means of the brilliance transfer defined as the ratio of the phase space density at the moderator and the sample positions. It follows from Liouville’s Theorem \cite{LandauLifshitz} that this density cannot increase, and thus the brilliance transfer is a value between zero and unity, where the latter is an indication of perfect transmission \cite{KlenoPhD,kleno2012}.

To compare our results to those found by Zendler {\em et al}, we also monitor the relative neutron transmission. This is calculated by normalizing the flux on the sample to that of a system with an ideal reference guide. Both of these methods of evaluation have the advantage of being normalized, thereby being independent of neutron intensity, wavelength etc. 

The results presented in this work have been simulated using the Monte Carlo ray tracing software package McStas version 2.5 \cite{Lefmann1999,manual,Willendrup2020}. By convention, McStas uses a right handed coordinate system with the $y$-axis in the vertical direction, and the flight path along the $z$-axis.

\subsection{Misalignment models}

As mentioned above, we implement two types of misalignment.
The first is the misalignment that arises from displacement between guide pieces during assembly, which is assumed to be random, and is implemented by displacing the beginning of each guide piece on both the $x$- and $y$-axis from its nominal position. The displacement is normally distributed with a standard deviation of $\sigma = 50 ~\upmu$m, as reported by one guide manufacturer \cite{snag}. Since the guide pieces are mounted in a common vacuum housing, the end of one guide piece is at the same position as the beginning of the next. Hence, the random error is cumulative, so that the total random displacement can add up to more than $50 ~\upmu$m.

\begin{figure}[b] 
        \includegraphics[width=9cm]{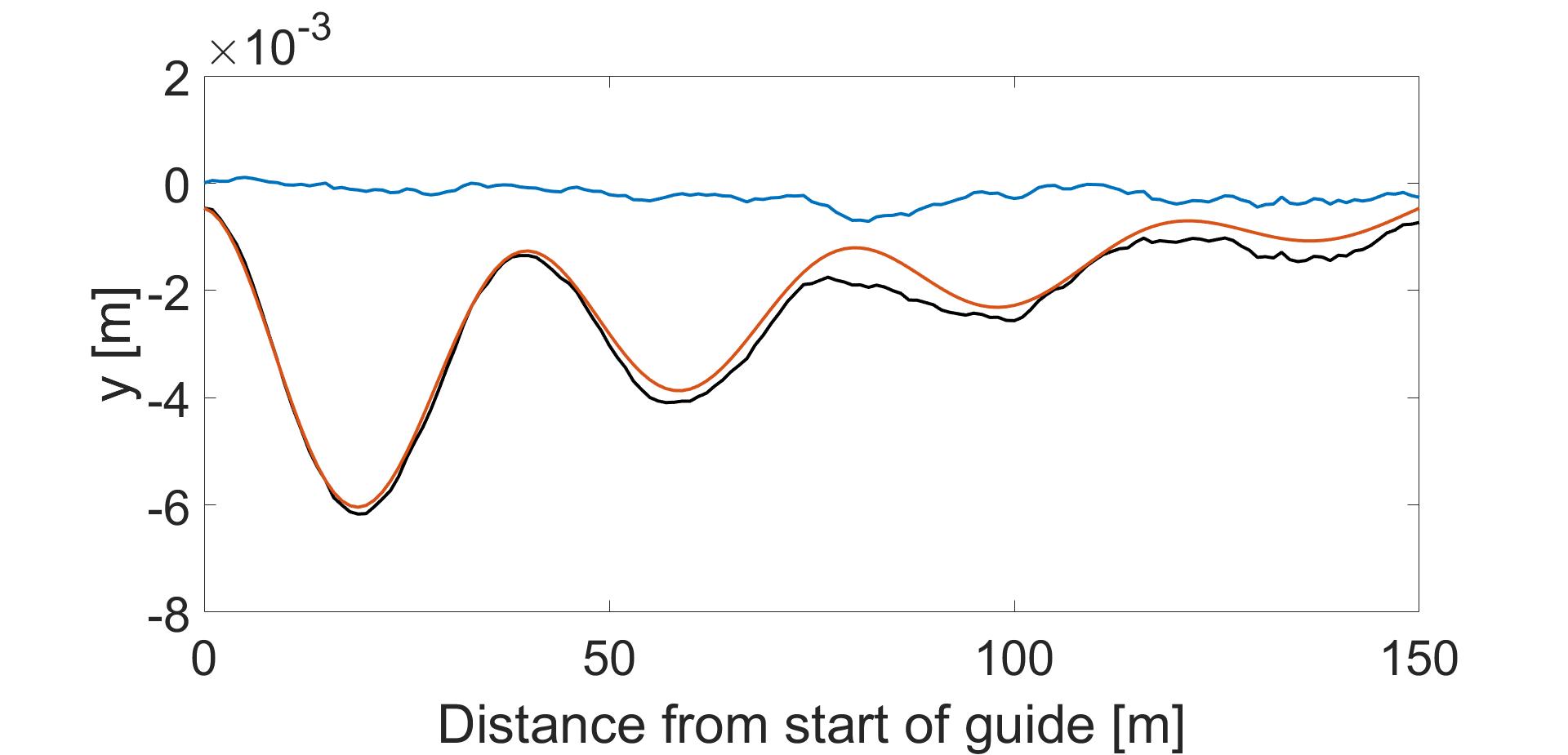}
        \caption{Model of the expected misalignment due to 6~mm maximal ground movement and random misalignment for a $150$~m guide. Blue line is the accumulated random misalignment, dark orange is the systematic ground movement given in eq.~(\ref{eq:deltay}). Black line is the total displacement of the guide.
        }
        \label{ground}
\end{figure}

The other type of misalignment is the systematic ground movement. To facilitate comparison, we adapt the model from  Zendler {\em et al}. In this model, misalignment appears only vertically (the $y$-direction), and is approximated by:

\begin{equation} \label{eq:deltay}
y = -A + A \cos \left( 2 \pi ~\frac{z}{z_p} \right)  e^{-Bz} + \frac{A}{L} \left( z-\frac{z_p }{2} \right) ,
\end{equation}
where $y$ is the vertical displacement, $A$ is the amplitude, and $L$ is the length of the guide. The other constants are chosen such that one period length is $40$ m, the oscillation amplitude is halved after each period, and the cosine function intersects the z-axis at $z = L$.  This model is based on experience from the Spallation Neutron Source (SNS), USA, adapted to the ESS lay-out \cite{Zendler2016}, and has been simulated for maximal displacements $y_{max}$ of 0, 1.5, 3, 6 and 12~mm. Figure \ref{ground} shows a plot of equation (\ref{eq:deltay}) for a maximum displacement of 6~mm, as this has been deemed most likely at ESS \cite{Zendler2016}, along with the accumulated random- and total displacements.

\begin{figure}[h] 
        \includegraphics[width=7cm]{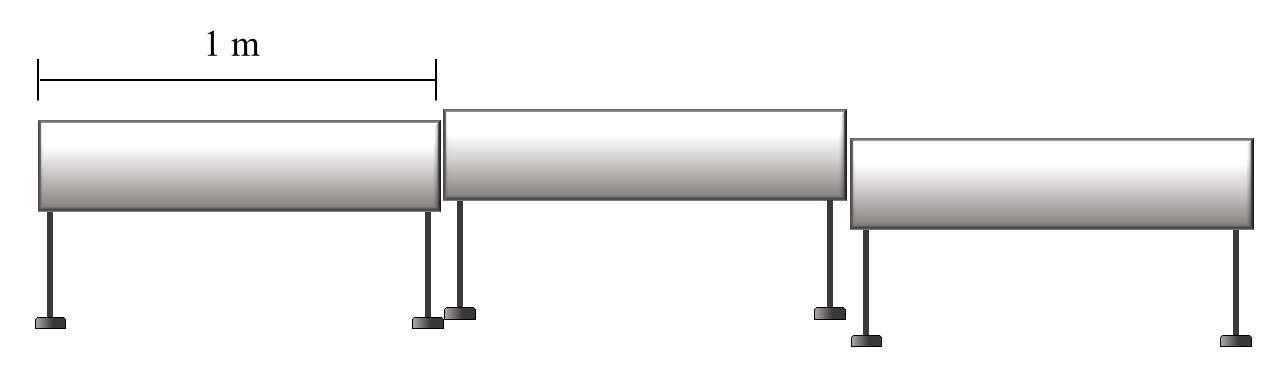}
        \caption{Model 1: Horizontal guide pieces of 1~m length, shifted in parallel with respect to one another in the fashion deviced by Zendler {\em et al.} \protect\cite{Zendler2016}.}
        \label{notilt}
\end{figure}

\begin{figure}[h] 
        \includegraphics[width=7cm]{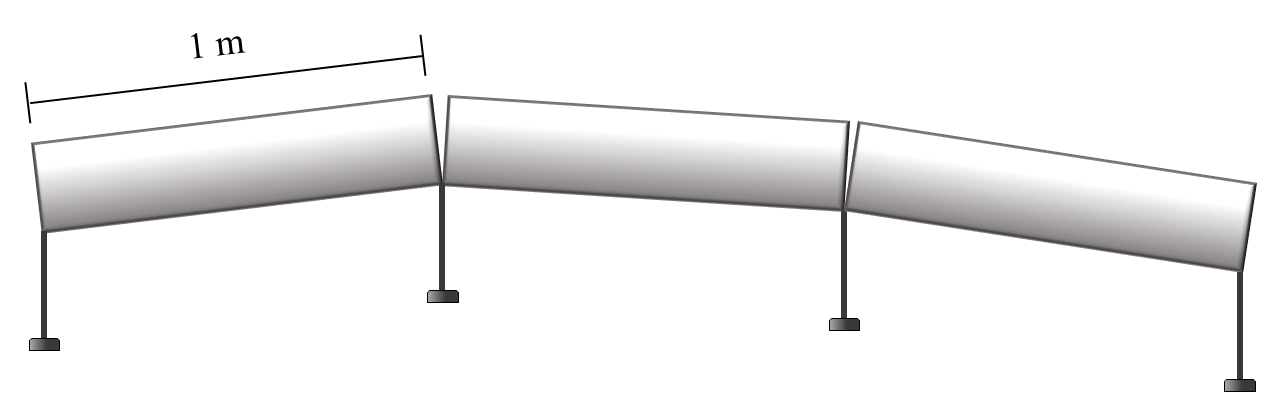}
        \caption{Model 2: Tilted guide pieces of 1~m length, where the ends of adjacent guide pieces are mounted on the same pillar.}
        \label{tilt}
\end{figure}

\begin{figure}[h] 
        \includegraphics[width=7cm]{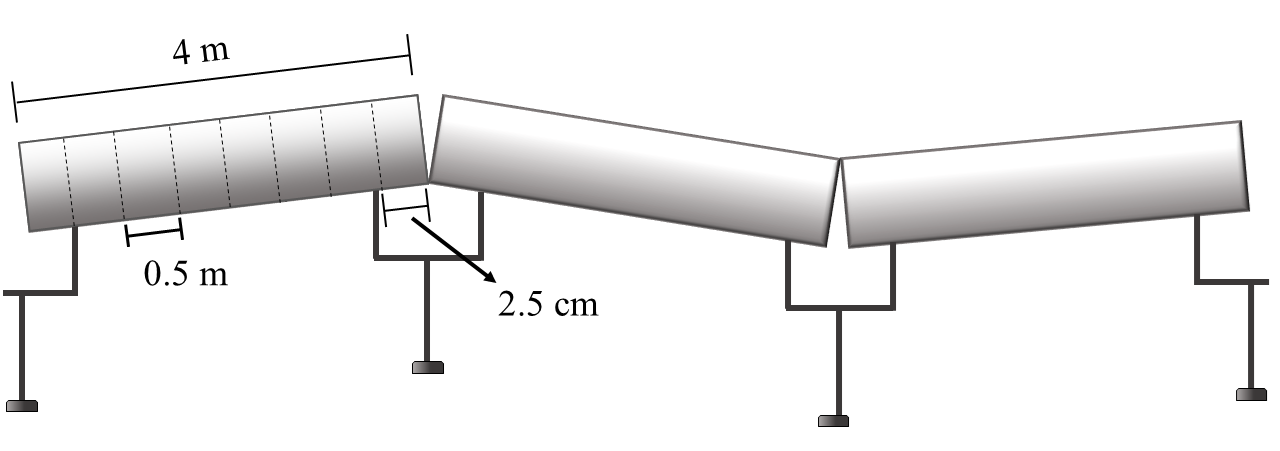}
        \caption{Model 3: The construction-realistic guide model. 8 guide pieces of length 0.5 m are placed in a form and rest upon pillars. The resting point, referred to as the pivot point is 2.5 cm from the edge of the guide.}
        \label{snag-guide}
\end{figure}

In our work, the two types of misalignment are implemented on 150 m long straight guides using three different models with progressively increasing complexity. The three models are outlined in figures \ref{notilt}, \ref{tilt} and \ref{snag-guide}. The figures are exaggerated for clarity and are not to scale. 

\begin{itemize}
\item \textbf{Model 1:} 1~m long guide pieces are shifted along the $x$- and $y$-axes, with the aim of reproducing the results by Zendler {\em et al}. See figure~\ref{notilt}. 

\item \textbf{Model 2:} 1~m long guide pieces are tilted around their starting points, as if two pieces are mounted at their corners to the same pillar. This closes the gaps, making the model more realistic. See figure~\ref{tilt}.

\item \textbf{Model 3:} is based on information from one possible vendor, and is the one closest to construction realities. Eight 0.5~m long guide pieces are placed inside 4~m long, inflexible mounting forms. These are, in turn, supported by holders on pillars with mounting points 2.5~cm from their ends, and the mounting forms pivot around this point, as seen in figure \ref{snag-guide}. Random displacement is added for every~0.5 m piece inside the forms, while the systematic ground movements are only added for each mounting form, {\em i.e.}\ every 4~m. This results in tilted mounting forms with small gaps between them.

\end{itemize}
To illustrate the practical implications of this work, the final model of misalignment is also implemented in a simulation of the guide for the ESS spectrometer BIFROST. Since the length of guide pieces for BIFROST are not constant, the model is implemented by placing 8 guide pieces in each form, and so the distance between pillars varies slightly. This approximation is assumed to be sufficiently accurate to give representative results.

\begin{table*}[t]
\centering
\caption{Simulation specifications for models 1, 2 and 3 on the straight guide, and for model 3 on the BIFROST guide. For a more detailed overview of the geometry of the BIFROST guide see  figure \ref{Bifrost}.
.}
\label{params}
\begin{tabular}{|l|l|l|}
\hline
Simulation  parameter & Straight guide & BIFROST \\ \hline
Source component & Source\_Simple() & ESS\_butterfly()\\
Monitors & L\_monitor() and Divergence\_monitor() & Same as for straight guide\\
Wavelength band & $0-10$ \AA & Same as for straight guide\\
Guide length & $150$ m & $ \approx 165$ m\\
Guide cross section & $3 \times 3$ cm$^2$ & Varies\\
Distance from moderator to guide entry & $2$ m & Same as for straight guide\\
Distance from guide exit to sample & $0.5$ m & Same as for straight guide\\
$m$-value & 3 & Varies\\
$\alpha$ & 3 & Varies\\
$q_{c,Ni}$ & $0.0219$ \AA$^{-1}$ & Varies\\
Standard deviation $\sigma$ & $0 ~ \upmu$m , $50 ~ \upmu$m & Same as for straight guide\\
Systematic amplitude $A$ & $0$, $0.00088$, $0.00176$, $0.00351$, $0.00703$ & 0, 0.00351, 0.00703\\
Maximal ground movement $y_{\rm max}$ & $0$ mm,  $1.5$ mm, $3$ mm, $6$ mm, $12$ mm &  0 mm, 6 mm, 12 mm\\ 
Systematic dampening factor $B$ & $0.017$ m $^{-1}$ & Same as for straight guide\\
Systematic period $z_p $  & $40$ m & Same as for straight guide\\ 
Small distance between guide pieces &$10 ~ \upmu$m  & Same as for straight guide\\
\hline
\end{tabular}
\end{table*}

\section{Simulation results}
A series of simulations have been performed in this work for which parameters are listed in table \ref{params}. Due to numerical instabilities, in the McStas component \textit{Guide\_gravity()}, for very small tilting angles, all simulations have been performed without gravitation. We still believe our simulations to be accurate, since experience has shown that the vertical shift of the beam in phase space, caused by gravitation, is counteracted by the bottom mirrors of the neutron guide.

\begin{figure}[t] 
        \includegraphics[width=9cm]{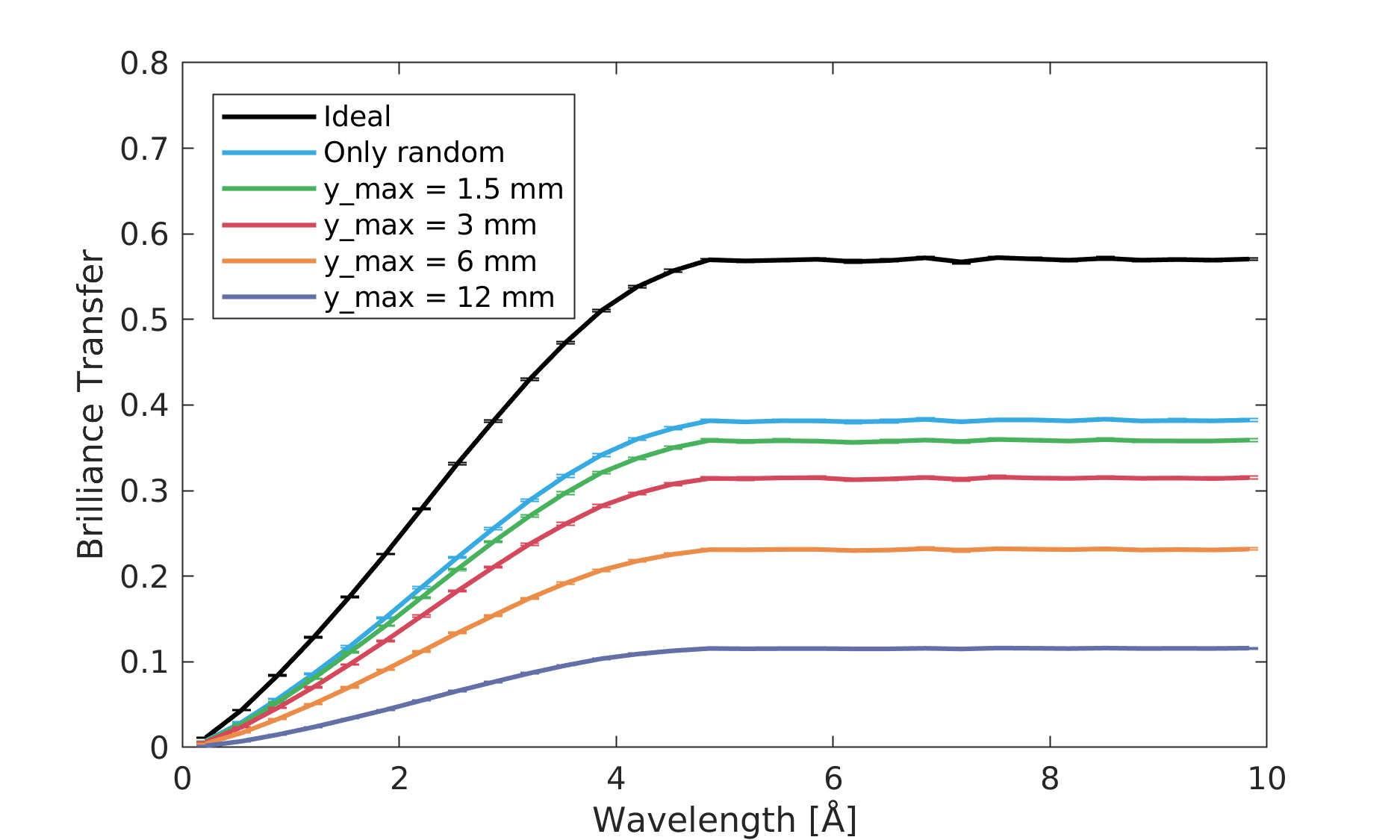}
        \caption{Brilliance transfer as a function of wavelength for a straight guide with displacement model 1.
        }
        \label{model1}
\end{figure}

\begin{figure}[t] 
        \includegraphics[width=9cm]{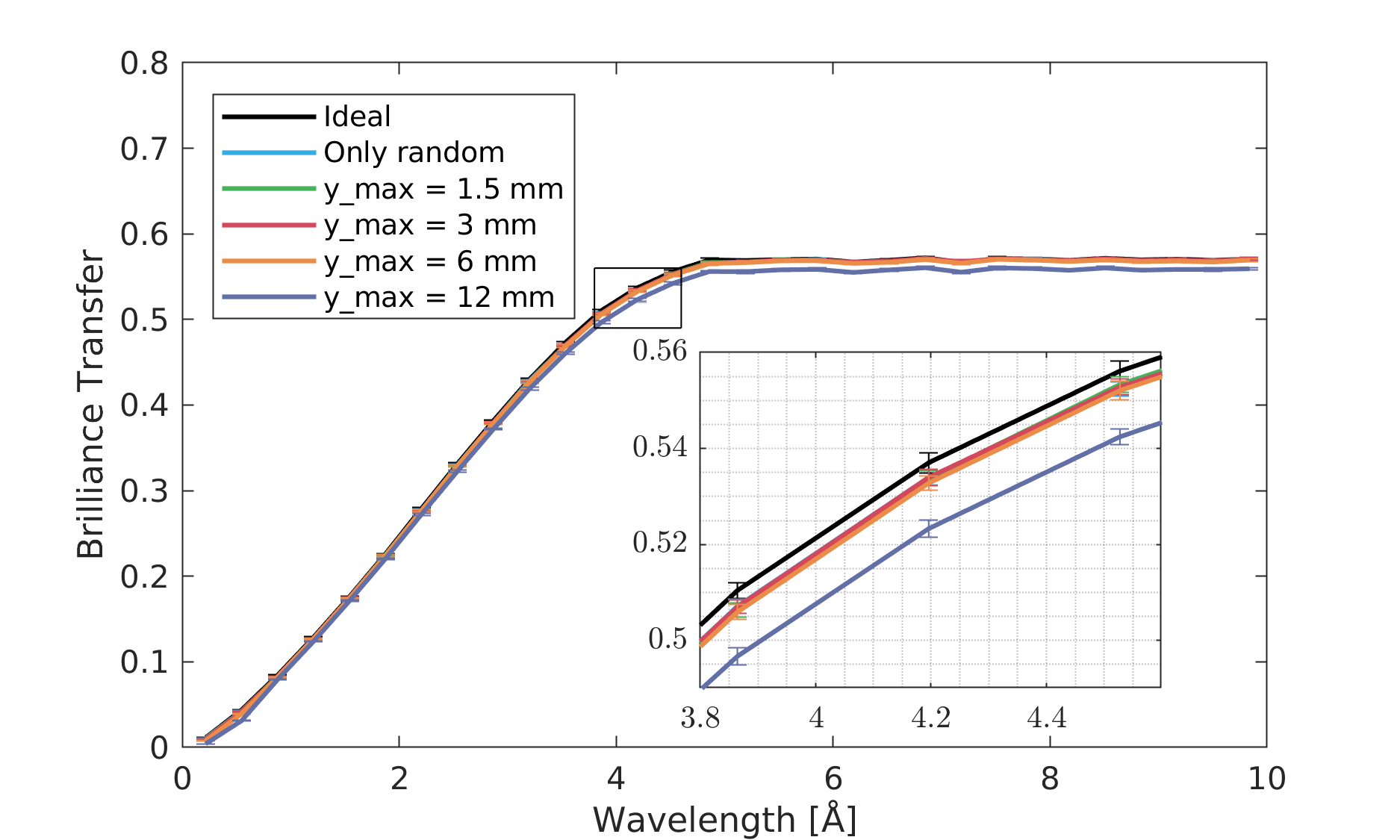}
        \caption{Brilliance transfer as a function of wavelength for a straight guide under displacement model 2. 
        }
        \label{model2}
\end{figure}

\begin{figure}[t] 
        \includegraphics[width=9cm]{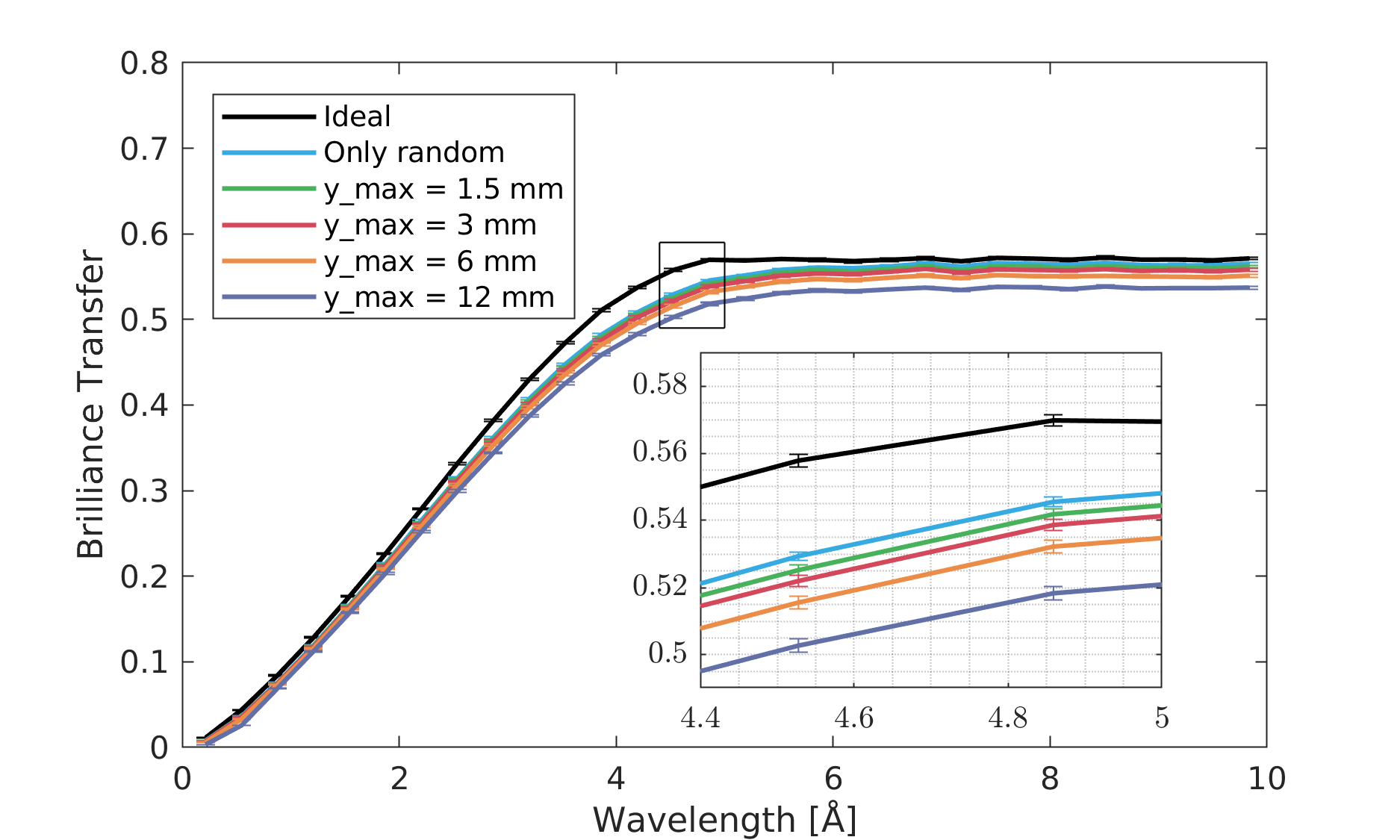}
        \caption{Brilliance transfer as a function of wavelength for a straight guide under displacement model 3, with 4 m guide sections and 0.5 m guide pieces.
        }
        \label{model3}
\end{figure}

Each system has been simulated multiple times, so as to average over both the Monte Carlo simulation fluctuations and variations from random changes of misalignment. In figures \ref{model1}, \ref{model2}, \ref{model3} and \ref{BIFROSTdata} the simulations have been run 30 times with $2 \times 10^8$ neutron rays. In figures \ref{BT/stand} and \ref{BT/y} each of the 20 data points represent 20 simulations with each $10^8$ neutron rays. All error bars are given as the standard deviation of the sampled data.

For table \ref{meanreltrans} the data has been simulated 25 times and errors tabulated are the estimated error on the results given, \textit{i.e.}\ the standard deviation of the mean.

Throughout this work data analysis and plotting have been performed using MATLAB \cite{MATLAB} and the add-on package iFit \cite{Farhi2014,iFit2}.

\begin{table*}[t]
\centering
\caption{Mean relative transmission given in percentages for $150$ m long straight guide from \cite{Zendler2016}, as well as for our simulations. The simulation of model 1 replicates the results of the previous study quite well. }
\label{meanreltrans}
\begin{tabular}{|p{2.2cm}|p{2.2cm}|p{2.2cm}|p{2.2cm}|p{2.2cm}|}
\hline
$y_{max}$ & $1.5$ mm & $3$ mm & $6$ mm & $12$ mm \\ \hline
Zendler {\em et al.} & $63.3 \pm 2.2\%$ & $54.0 \pm 1.3\%$ & $40.2 \pm 0.5 \%$ & $20.8 \pm 0.3 \%$ \\ \hline
Model 1 & $62.90 \pm 0.04\%$ & $55.26 \pm 0.04\%$ & $40.73 \pm 0.03\%$ & $20.80 \pm 0.02\%$\\ \hline
Model 2 & $99.63 \pm 0.01\%$ & $99.64 \pm 0.01\%$ & $99.45 \pm 0.01\%$ & $97.61 \pm 0.01\%$\\ \hline
Model 3 & $96.53 \pm 0.02\%$ & $95.94 \pm 0.02\%$ & $94.68 \pm 0.02\%$ & $92.18 \pm 0.02\%$\\ \hline
\end{tabular}
\end{table*}

\subsection{Ground movement effects on a straight guide}
For measuring relative transmission, we use wavelength-sensitive monitors detecting all neutrons reaching the sample (maximal divergence of $3 \degree$). A simulation  with a straight guide (zero misalignment) is used as the reference guide.
The relative transmission is calculated by dividing the output of the wavelength monitor by the reference guide results. 
The mean relative transmission over the entire wavelength interval is shown in table \ref{meanreltrans} along with the results from Zendler {\em et al.}
We see that all our results for model 1 lie within the error margin of the Zendler {\em et al.}\ data. 

To calculate the brilliance transfer, we use wavelength-sensitive monitors with the same area as the guide, sensitive only within a divergence of $\pm 0.5 \degree$, placed at the source and sample positions, respectively. Simulations have been run for a guide with only random misalignment ($y_{\rm max} = 0, \sigma =50 ~ \upmu $m), as well as the four other values of $y_{\rm max}$. The ratio of the monitor outputs for the three geometries are plotted in figures \ref{model1}, \ref{model2} and \ref{model3} for model 1, 2 and 3, respectively, and the black graph for the 'ideal' guide is the same in the three plots.

\begin{figure}[t] 
        \includegraphics[width=8.5cm]{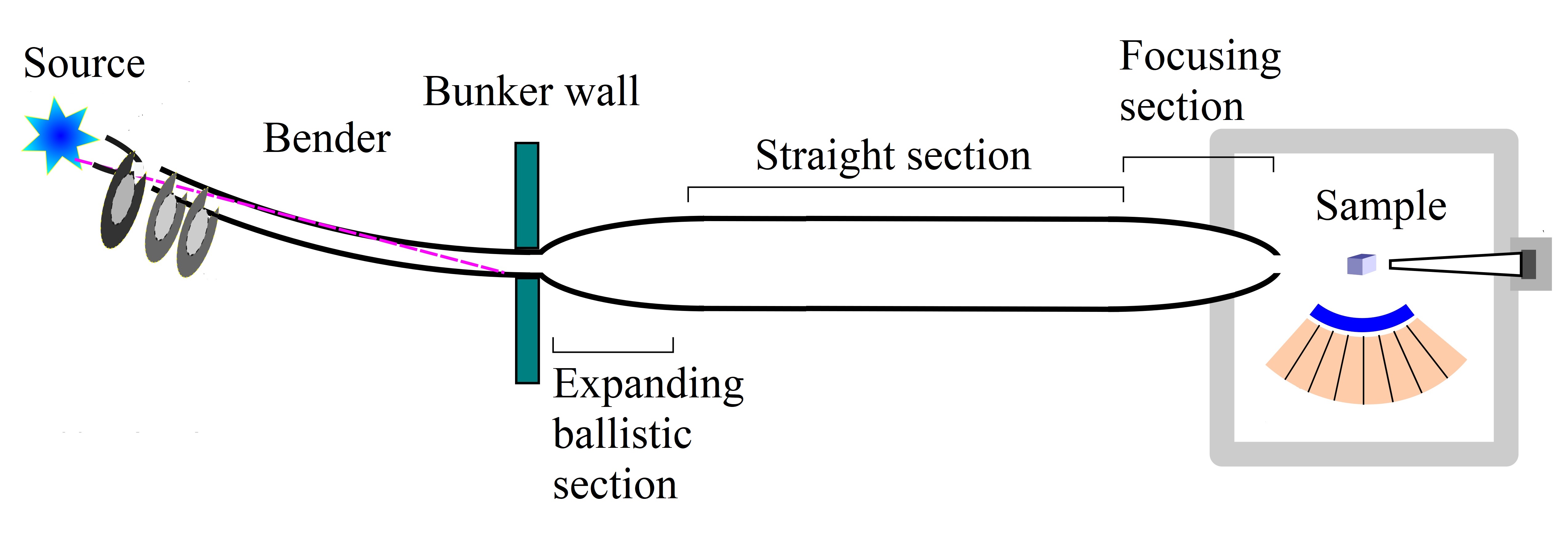}
        \caption{Rough outline of the BIFROST guide geometry. The figure is not to scale. From Ref.~\cite{geometry}. }
        \label{Bifrost}
\end{figure}

\subsection{Ground movement effects on the BIFROST guide}

The BIFROST guide consists of a  half-ellipse feeder, a curved guide and a ballistic section with an expanding- and a focusing ellipse at each end. The sample area of interest for BIFROST is $10 \times 10 ~\textrm{mm} ^2$ with a maximal divergence of $\pm 0.75 \degree \times \pm 0.75 \degree$. A detailed description of guide dimensions and coating values can be found in 
Ref.~\onlinecite{Olsen2020}, and a simple sketch of the guide geometry can be seen in figure \ref{Bifrost}. 

The model 3 misalignments have been added to the bender-, expanding-, straight- and focusing parts of the guide. Since truly curved guide mirrors are more expensive than straight ones, it is cost effective to approximate the geometry by using straight guide pieces, and has only minor impact on the total performance \cite{Froystad2020}. Therefore only straight guide pieces have been used for this simulation.

\begin{figure}[] 
        \includegraphics[width=9cm]{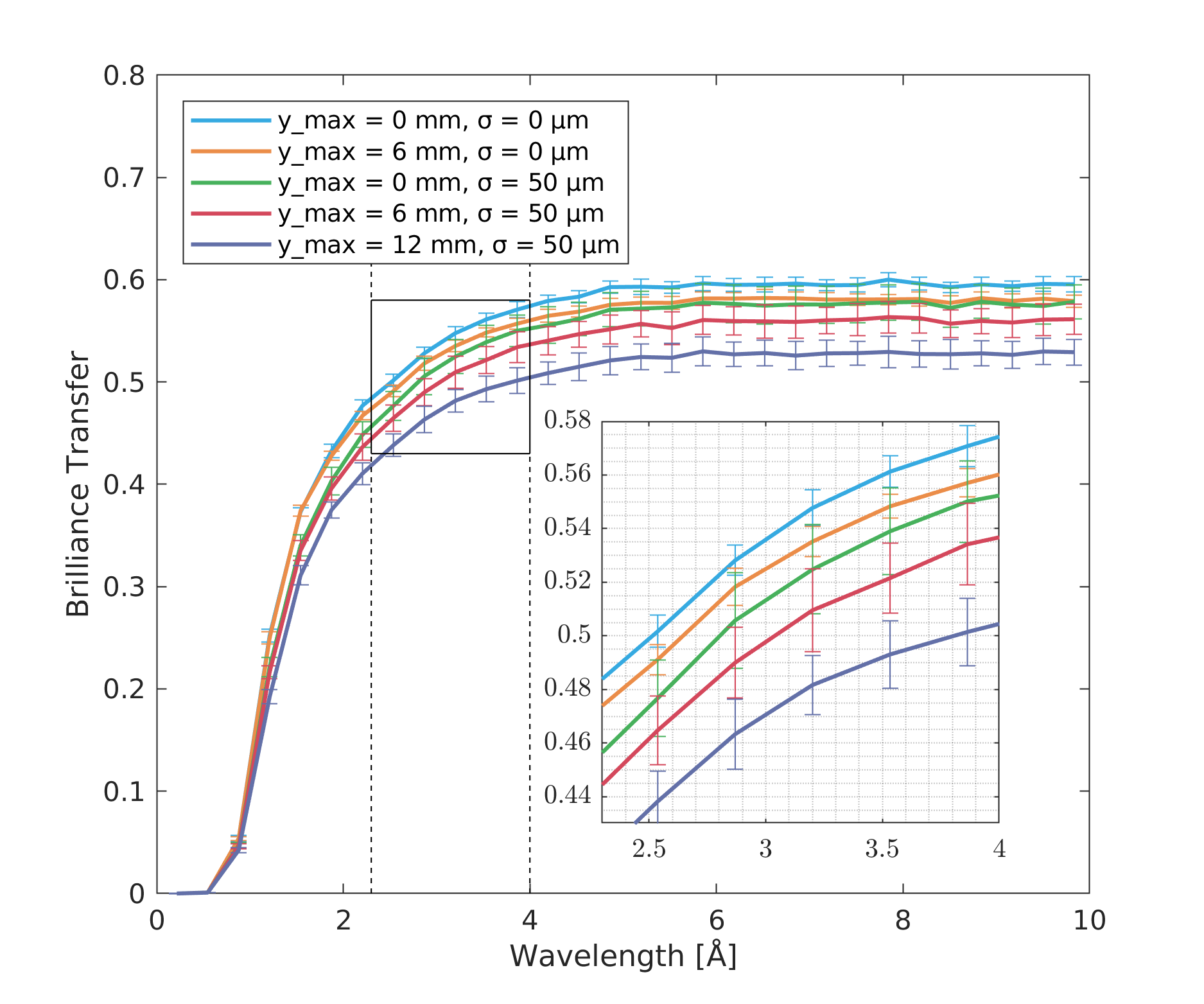}
        \caption{Brilliance transfer as a function of wavelength for the BIFROST guide with model 3. The blue curve is the ideal guide. The dotted, vertical lines show the wavelength band 2.3 - 4.0 Å.
        }
        \label{BIFROSTdata}
\end{figure}

A monitor is placed at the sample position with the relevant specifications mentioned above and, as the brilliance reference, an identical monitor is placed on the surface of the ESS butterfly moderator, centered on the spot on the cold part where the guide is focused. The wavelength-dependent brilliance transfer is displayed in figure~\ref{BIFROSTdata}. Here, the blue curve is the benchmark simulation with no ground movement added, and red curve represents the most likely scenario with  maximum ground movement of $y_{\rm max} = 6$~mm. We note that the total reduction due to both ground movements and random misalignment is minor, approximately 0.04 in brilliance transfer, corresponding to a 7\% loss.

\begin{figure}[] 
        \includegraphics[width=9cm]{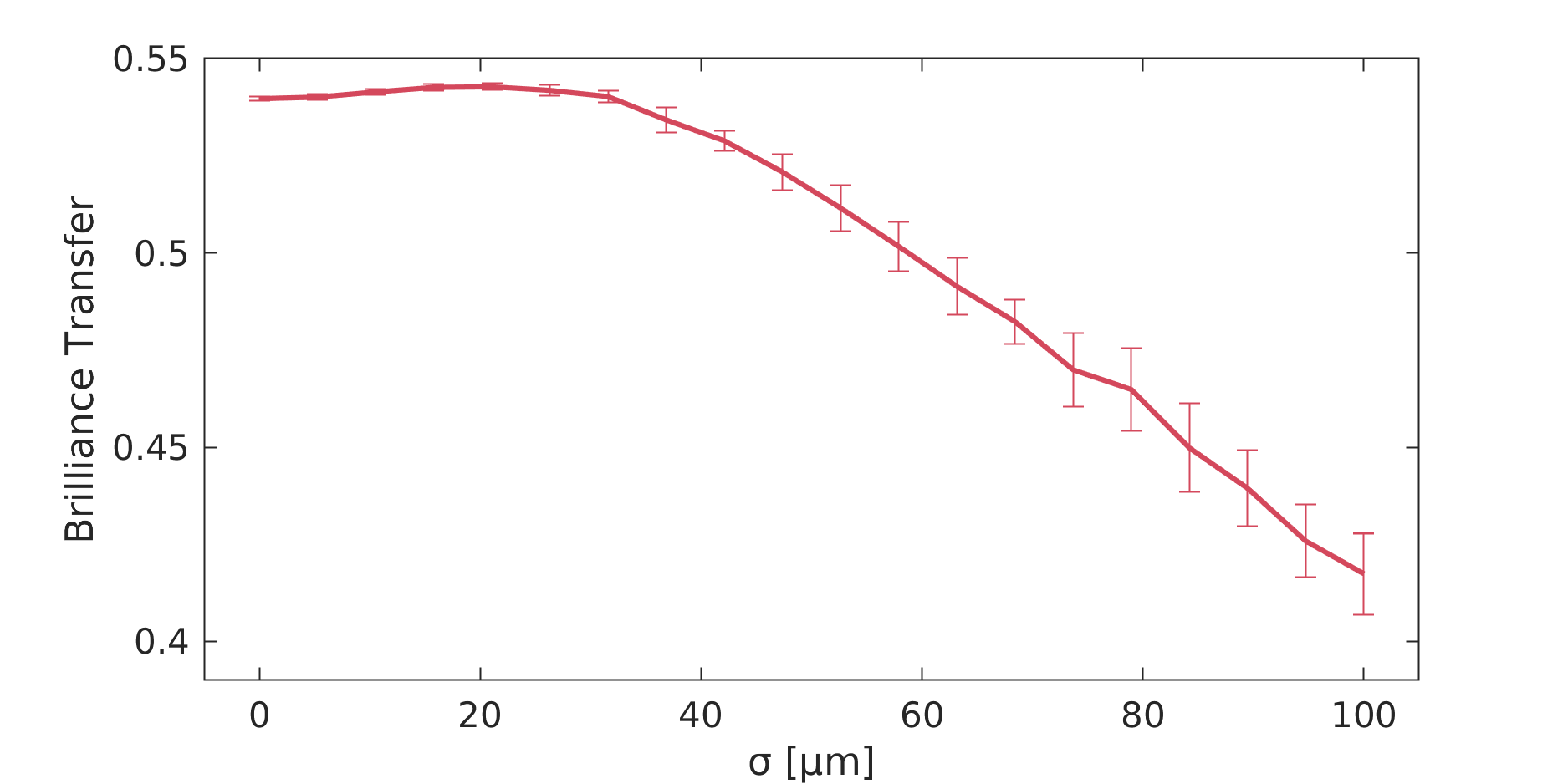}
        \caption{Brilliance transfer as a function of the random misalignment. Brilliance transfer is average in the range 2.3 - 4.0 Å.}
        \label{BT/stand}
\end{figure}

\begin{figure}[] 
        \includegraphics[width=9cm]{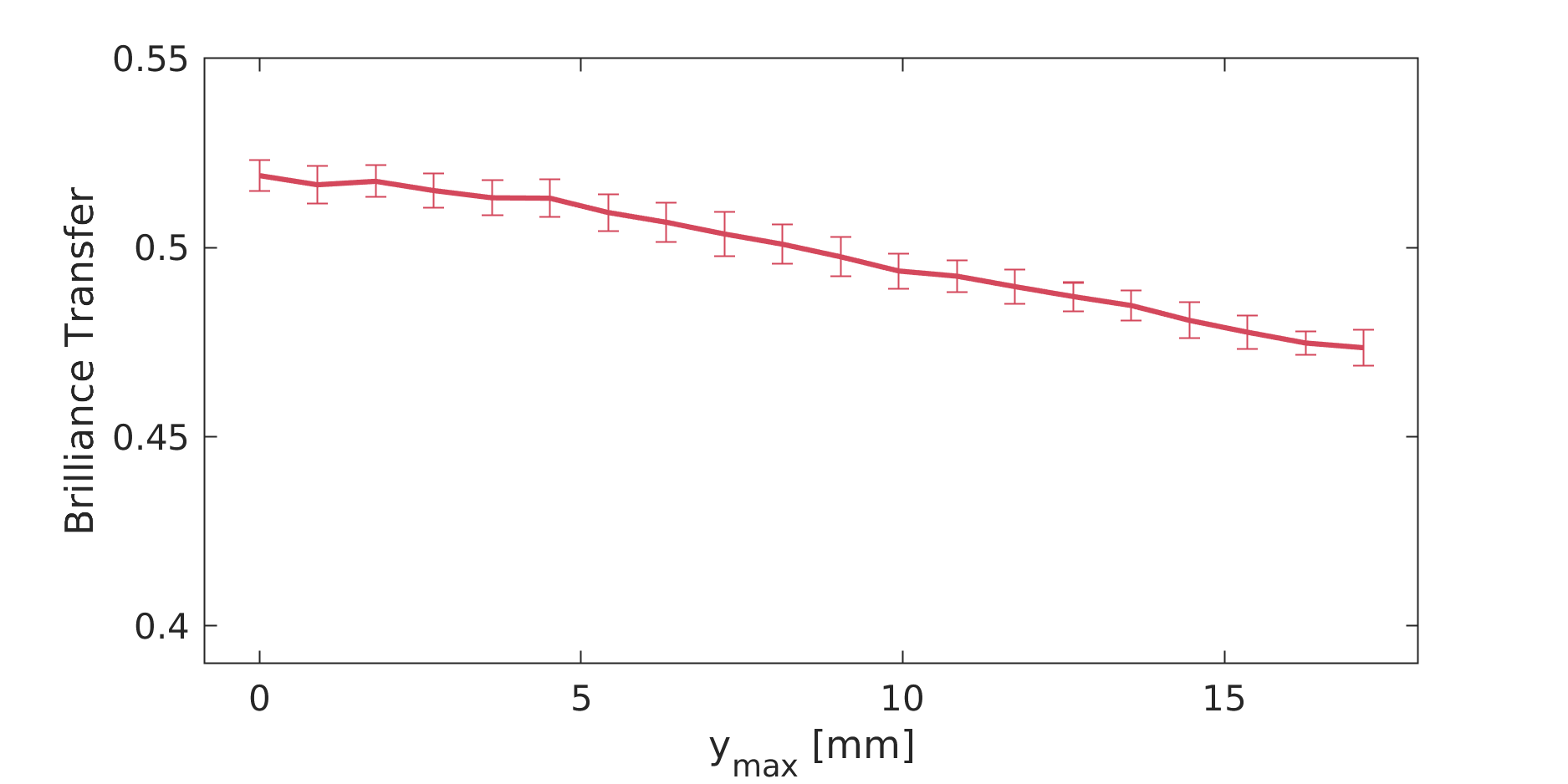} 
        \caption{Brilliance transfer as a function of the ground movements. Simulated with a random misalignment of $\sigma = 50 ~ \upmu$m. Brilliance transfer is average in the range 2.3 - 4.0 Å. }
        \label{BT/y}
\end{figure}

The effect on brilliance transfer of random misalignment is shown in figure \ref{BT/stand}. Small random misalignments, has hardly any effect, but for $\sigma > 30 \; \upmu$m, the brilliance transfer starts decreasing linearly with $\sigma$. In addition, the standard deviations of the brilliance transfer increase due to enhanced random effects on the guide geometry. The effect of ground movement amplitude is displayed in figure \ref{BT/y}. Here, the brilliance transfer shows a slow, linear decrease over the whole interval, with a constant standard deviation. 


\begin{figure}[t] 
        \includegraphics[width=8.5cm]{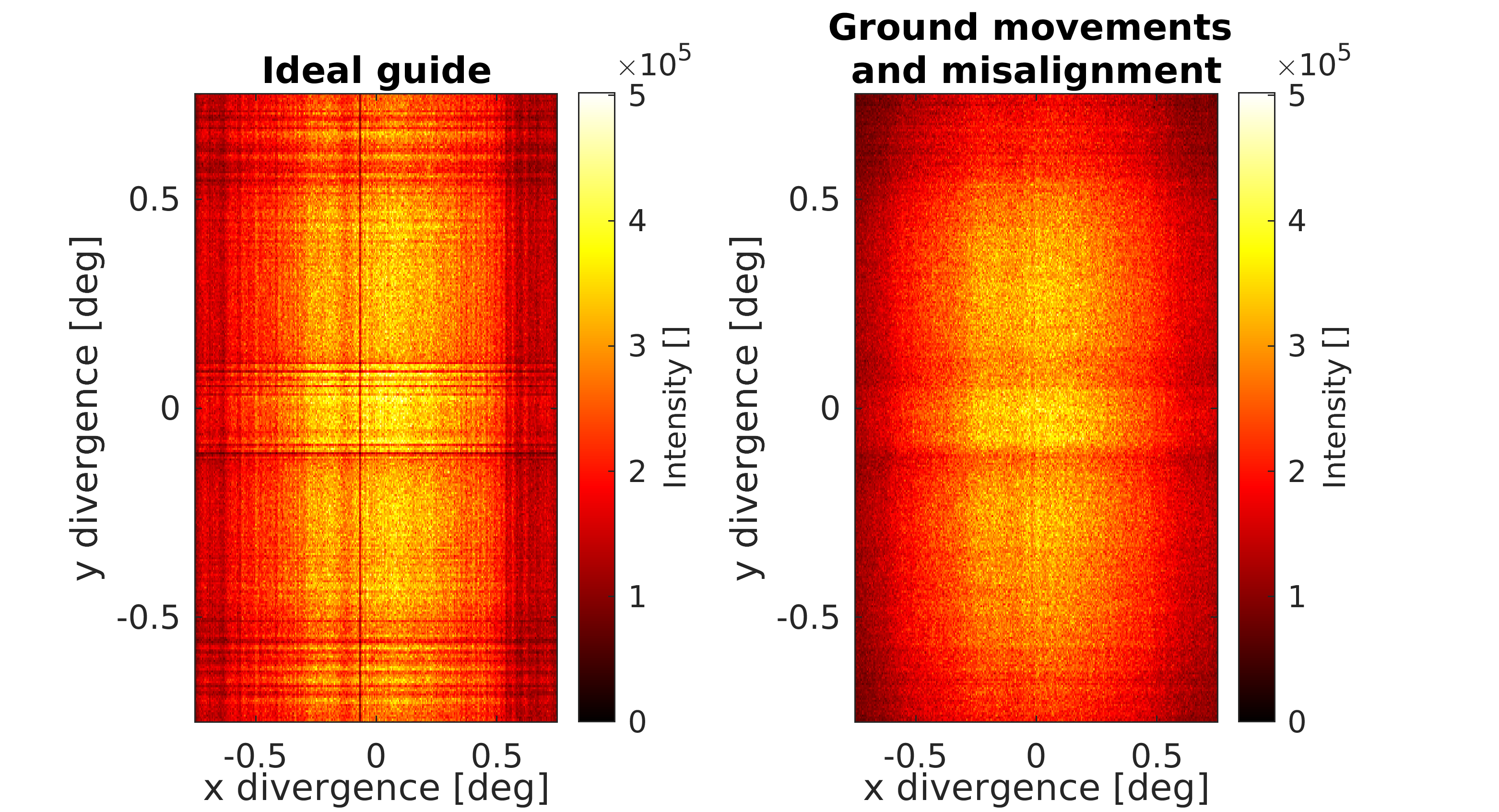}
        \caption{Simulated divergence profile at the BIFROST sample position, integrated over the spectrum of the cold ESS moderator. Left panel shows results for the ideal guide, and right panel for $\sigma = 50 ~ \upmu$m and ground movements of 6 mm. In the color scale, yellow represents the highest intensity.}
        \label{sample}
\end{figure}

In figure \ref{sample}, the output of the divergence monitor at the sample position is plotted for two typical single guide settings: One with misalignment and ground movements and one without these effects. The distinct vertical and horizontal lines that come from the 0.5 meter segmentation of the focusing ellipse are almost smeared out as an effect of the misaligments.

\section{Discussion}
\subsection{Wavelength dependence of the brilliance transfer}

In the study of the straight guide, for the 3 models as well as for the ideal guide (figures \ref{model1}, \ref{model2} and \ref{model3}) we see that the brilliance transfer tends towards zero for small wavelengths. This is expected, since the critical scattering angle is proportional to the wavelength. In a model where the reflectivity value was either zero or unity, the wavelength dependence of $\theta_{\rm c}$ would lead to a brilliance transfer that depends on $\lambda^2$, since the relevant divergence phase space is two-dimensional. However, after a (possibly) quadratic start at very low $\lambda$-values, we observe an almost linear increase up to $\lambda = 4.5$~\AA. This flatter increase of brilliance transfer is caused by the reflectivity values being smaller than unity for a large part of the scattering angles, combined with the fact that larger scattering angles give rise to larger number of reflections (and thus reflection losses) in the straight guides.

For the BIFROST guide the quadratic tendency is seen for a larger span of wavelengths, up to around 1.0~\AA , and the brilliance transfer reaches its maximum value already from around 2.5~\AA . This is understood by that fact that the elliptic geometry decreases the amount of reflections in the guide \cite{kleno2012} combined with  higher $m$-values at critical points along the guide. All in all this explains that the geometry of the guide combined with a targeted distribution of $m$-values allows more neutrons in the wavelength range 1.5-4~\AA\ compared to the straight guides. 

\subsection{Reproducibility of simulation results}
As we discuss below, our results deviate in some points from those of Ref.~\cite{Zendler2016}. It is therefore prudent to start with reconciling our simulation methods. The previous work used the simulation package VITESS, while we employ McStas. This could be a source of discrepancy. However, the two packages (as well as the similar package RESTRAX/SIMRES) have been extensively benchmarked between each other \cite{kleno2012}, so the use of a different package is not supposed to cause differences.

Our initial simulations of model 1, displayed in Table~\ref{meanreltrans}, were performed exactly for the purpose of testing the possible difference between programs. Since all of the results fall within the (fairly tight) margins of error of the previous data, we conclude that we have successfully replicated the simulation and that any difference between simulation packages is insubstantial. 

Since the simulations performed by Zendler {\em et al.} included gravitation this also confirms that gravitation does not have a significant impact on the transport properties of the guide.

\subsection{The effect of tilting: Model 2}

Our simulation of model 2 shows a clear improvement of relative transmission (see table \ref{meanreltrans}) and brilliance transfer (see figures \ref{model1} and \ref{model2}) compared to model 1. In fact, misalignment seems to have almost no effect on the neutron transport, indicating that the source of losses is not the tilting of guide pieces, but rather the gaps between them.

Since misalignment, and specifically guide gaps, has such a great effect on the resulting transport properties, using a construction realistic geometry is paramount in order to estimate the true beam losses in a particular guide.


\subsection{The construction-realistic scenario: Model 3}
In figure \ref{model3}, we see that the transport properties of the model 3 guide are worse than for model 2 but far better than model 1. The relative transmission indicates a loss of $ \sim 5 \%$ of the neutron flux, for the expected $y_{\rm max} = 6$~mm guide, compared to  $ \sim 60 \%$ for model 1. This is to be expected, since the construction realistic model 3 has much less severe gaps than for model 1, indicating that long guides are not as sensitive to misalignment as indicated in Ref.~\cite{Zendler2016}.

\subsection{The effect of ground movements on the BIFROST guide}

As can be seen from figure \ref{BIFROSTdata}, the effect of realistic ground movements on the BIFROST neutron guide does have some impact on the transport properties of the guide. Compared to the simulation with no misalignment, the relative loss of neutrons in the relevant wavelength- and divergence interval, caused by $y_{max} =6$ mm is $7.0(5) \%$, and for $y_{max} =12$~mm the loss increases to $12.2(4) \%$. This loss is slightly greater than what was found for the model 3 straight guide. Hence, it seems that the more complex BIFROST guide geometry, which is not everywhere elliptical in shape, is slightly more sensitive to misalignment. However, the guide still has a high absolute brilliance transfer, above 0.5 for most wavelengths, in the desired phase space.

The effect of random misalignment of $\sigma = 50 \; \upmu$m seems to be of greater magnitude than that of the systematic ground movements of $y_{\rm max} = 6$~mm, when comparing figure \ref{BT/stand} and \ref{BT/y}. Reducing the random misalignment to around $30 ~ \upmu$m could be strongly beneficial to increase the brilliance transfer, but not much brilliance transfer is expected to be gained by going below this misalignment value.

Furthermore, figure \ref{sample} shows that the misalignments seem to have the beneficial side effect of reducing both the horizontal and vertical lines over lower intensities in the divergence space. These lines originate from the piecewise tapering of the focusing ellipse \cite{Froystad2020}. 

It is clear that the results presented in this work depend upon the ground displacement model given in eq.~(\ref{eq:deltay}). It is unlikely that the ground under ESS will sink in exactly this pattern. Nevertheless, we do not expect a difference in the ground movement model to give significantly different results, as we have shown that it is in fact the random misalignments that give rise to most of the losses.

\subsection{Mitigation}

Based on simulations performed with the model 1 geometry it has been suggested \cite{Zendler2015,Zendler2016} that BIFROST and other long instruments at ESS could make use of vertical overillumination at strategic positions of the guide, as a means of mitigating the loss of neutrons caused by misalignment. Overillumination is achieved by decreasing the vertical cross section of guide pieces following sections, at which there is expected to be significant beam loss, so as to prevent gaps in the phase space. That is, the height of the first guide pieces will be made higher that originally planned for. Although this method can be efficient in preventing large losses of flux, in particular within model 1, it does introduce losses of its own. In particular this is the case for the ESS, since the short (3~cm) moderator already without overillumination does not allow for a full vertical illumination of the first guide pieces \cite{Holm2019}. For taller guide pieces, this effect will clearly worsen. In contrast, our present investigation indicates that the loss of flux due to ground movements will be relatively small. For this reason, we do not believe that mitigation by overillumination is beneficial for long guides within the construction-realistic model 3.

It is worth noting, that since higher $m$-value mirrors are more expensive, the coating for the BIFROST guide has been optimized for price versus performance \cite{Olsen2020}, and thus the $m$-values are everywhere at a minimum level. Therefore, a possible mitigation method could be to overcoat the guide, {\em i.e.}\ make use of neutron mirrors with higher $m$-values. This would in result in a higher neutron flux of higher divergence neutrons. However, most likely these neutrons would have so high divergence as to be placed outside the desired phase space, thereby not adding to the brilliance value. However, a detailed study would be necessary in order to properly evaluate this mitigation method.

\section{Conclusion}
In this work a series of simulations have been performed with the goal of investigating the effect of misalignment on long neutron guides and specifically the effect on the BIFROST instrument currently under construction at ESS. 
We have implemented three models for misalignment on a 150 m straight guide and managed to replicate the results found in Ref.\ \onlinecite{Zendler2016} for their model, giving credence to our method. The most construction-realistic model has been used to simulate the full BIFROST guide.

Our results suggest that the neutron losses come primarily from the gaps prompted by vertical misalignment and much less from the tilting of guide pieces. In addition, the BIFROST simulations have shown that the random misalignment associated with the installment of guide pieces, can potentially be a greater source of loss than the systematic ground movements. However, in both cases, the expected losses are at an acceptable level.

We find that the effect of misalignment on long neutron guides is significant, but smaller than previously expected. For the 6~mm maximal ground movement expected at ESS, we have found a loss in neutron flux of $7.0(5)\% $ for the BIFROST guide, and we expect similar effects on other long instruments at ESS. This result suggests that it is likely not advantageous to mitigate misalignment losses by overillumination of long neutron guides. 

\section*{Acknowledgments}
We thank Michael Schneider from Swiss Neutronics A.G.\ for valuable discussions on the engineering details of neutron guides.

\bibliography{guidebib}